\documentclass[12pt]{iopart}

\begin{document}

\title{A conserved parity operator}
\author{Mark J Hadley}
\address{Department of Physics, University of Warwick, Coventry
CV4~7AL, UK\\ email: Mark.Hadley@warwick.ac.uk}

\begin{abstract}
The symmetry of Nature under a Space Inversion is described by a
Parity operator. Contrary to popular belief, the Parity operator
is not unique. The choice of the Parity operator requires several
arbitrary decisions to be made. It is shown that alternative,
equally plausible, choices leads to the definition of a Parity
operator that is conserved by the Weak Interactions. The operator
commonly known as $CP$ is a more appropriate choice for a Parity operator.
\end{abstract}

\vspace{20mm}
{\footnotesize Pacs 11.30.Er, Charge conjugation, parity, time
reversal, and other discrete symmetries}

\maketitle

\section{Introduction}
All the classical laws of Nature are symmetric under a Parity
operation. However, in particle physics the weak interactions
seemed to show some asymmetry between left and right. The
experiments, and the theory that described the interactions, were
taken as clear evidence of an inexplicable asymmetry in Nature.
The ramifications of the experiments extend far beyond the weak
interactions and particle physics; they color our view of the
Universe and are widely referred to in textbooks that discuss the
nature of space and time.

This paper shows that parity violation, is simply a consequence of
the arbitrary assumptions that are made in the choice of Parity
operator.

Symmetry under space inversion is described by a Parity operator
$P$. Under space inversion we have by definition:
\begin{equation}
\begin{array}{lccr}
\left.
  \begin{array}{ccr}
  x & \rightarrow & -x \\
  y & \rightarrow & -y \\
  z & \rightarrow & -z \\
  \end{array}
  \right\}
   & \ \textbf{r} & \rightarrow & -\textbf{r}\\
  {\begin{array}{ccr}
  t & \rightarrow & t\\
  \end{array}} &  &  & \\
\end{array}
 \end{equation}

It follows from the definition of $P$ that velocity and
acceleration vectors also change sign. Given a mathematical object
defined in terms of co-ordinates and derivatives, the transformation
properties can be determined
(eg angular velocity $\textbf{r}\times \dot{\textbf{r}}$ is unchanged). However this is not
sufficient to define  the parity operator because we do not know
the transformation properties of other entities that appear in the
equations of Physics. In most cases there is a choice of mathematical
representations and the most obvious choice is not necessarily the most appropriate.

\section{Electrodynamics}

Electrodynamics offers a neat illustration of the choices that
need to be made when defining a Parity operation - and the
importance of making the best choice. Choosing a parity operator is
equivalent to choosing mathematical representations for the physical quantities that appear in the equations.

Maxwell's equations have an Electric Field, \textbf{E}, and a
Magnetic field, \textbf{B}. A simple mathematical representation  would be to treat
them both as vectors. Both \textbf{E} and \textbf{B} would then
change sign under a parity operation. Maxwell's equation would not
be invariant under this parity operation. The Lorentz equation:
\begin{equation}
\label{eq:lorentz} \ddot{\textbf{r}} = q(\textbf{E} +
\dot{\textbf{r}} \times \textbf{B})/m
\end{equation}
Would be composed of a parity violating term $q \dot{\textbf{r}}
\times \textbf{B}/m$ and a parity conserving term $q \textbf{E}$.

Electrodynamics would be inherently lefthanded. A symmetrical
experimental arrangement such as an electron beam moving forward
in the $x$-direction and a magnetic field pointing in the
$z$-direction would result in electrons moving to the left. The
mirror image would be contrary to experimental observations.

The predictions of electromagnetism would be unchanged by the choice of
parity operator. Since a parity operation cannot be performed in the
laboratory, there could be no direct test of the transformation
properties of \textbf{E} and \textbf{B}. The most that experiment
could do is confirm the two different parts of the Lorentz
equation. Particle motions would still be correctly
described regardless of the transformation properties (under parity) of \textbf{E} and \textbf{B}.

Although the choice of operator is of no experimental
significance, it is certainly of philosophical interest for it
describes, an asymmetric Universe. But the choice of operator is
far more significant than just philosophical. The choice of
operator clarifies (or confuses) a theoretical understanding of
the Physics, which is vital for progress. It is particularly
important when relating electrodynamics to relativity.

Either \textbf{E} or \textbf{B}, or both, can be axial vectors
that do not change sign under a parity operation. If they are both
axial vectors then electrodynamics does not conserve Parity. But
if only \textbf{B} is an axial vector then the parity operator
transforms \textbf{E} to \textbf{-E} and leaves \textbf{B}
unchanged.  With this definition Electrodynamics conserves parity.

The different transformation properties of the Electric and
Magnetic fields seems puzzling and arbitrary. However if Maxwell's
equations are described using the Faraday two-form (antisymmetric
second rank tensor) the Magnetic field is seen as the space-space
components while the Electric field is the space-time components
of the same mathematical entity.
\begin{equation}\label{eq:F}
    \textbf{F} =
\left(%
\begin{array}{cccc}
     0 & - E_x  & - E_y  & - E_z \\
  E_x & 0    & B_z  & -B_y \\
  E_y & -B_z & 0    & B_x \\
  E_Z & B_y  & -B_x & 0   \\
\end{array}%
\right)
\end{equation}
The dual of the Faraday two form is the Maxwell two form defined
by $\textbf{M} = \textbf{*F}$. The dual interchanges the position
of the electric and magnetic fields. The source-free Maxwell's equations
can then be written in the most elegant form $d \textbf{F} = 0$
and $d \textbf{M} = 0$

The different nature of the electric and magnetic fields is also
evident if they are described using a four-vector potential $A_\mu$ with:
\begin{eqnarray}\label{eq:A}
  \textbf{E} &=& -\nabla A_t - \partial_t (A_x,A_y,A_z) \\
  \textbf{B} &=& \nabla \times (A_x,A_y,A_z)
\end{eqnarray}
From which it is clear that \textbf{E} is a vector and \textbf{B} is an axial vector.

The definition of the Parity operator assumed that mass, $m$, in
the Lorentz equation (\ref{eq:lorentz}) was unchanged. The charge
$q$ is unchanged but this is not a separate assumption because the
charge, $q$, in a volume can be defined from the surface integral
of electric field evaluated over a spherical surface that encloses
the volume:
\begin{equation}
\label{eq:charge}
\begin{array}{cccccccc}
 P:& q & = & \oint \textbf{E}.\hat{\textbf{n}}\  \rm{dS} &\rightarrow &\oint (\textbf{-E}).(-\hat{\textbf{n}})\  \rm{dS} &=& q
\end{array}
\end{equation}
A similar definition of magnetic charge $Q_m$ is:
\begin{equation}
\label{eq:qm}
\begin{array}{cccccccc}
  P:&Q_m & = & \oint \textbf{B}.\hat{\textbf{n}}\  \rm{dS} &\rightarrow &\oint \textbf{B}.(-\hat{\textbf{n}})\  \rm{dS} &=& -Q_m
\end{array}
\end{equation}
From which it is clear that a Parity operation will change the
sign of a magnetic charge. Under a reflection a North pole changes
to a South pole and vice versa. The intimate relation between charges and fields is
such that the Parity operator can equally well be defined through the effect
on the charges (magnetic and electric). Conventionally electric charge is
represented as a scalar quantity and the magnetic charge as pseudoscalar.

It is convention to choose the magnetic field to be an axial
vector, but it is well known~\cite{Jackson,MTW} that the equations
of electrodynamics are invariant under a duality rotation that
interchanges the role of the \textbf{E} and \textbf{B} fields.
With this unconventional representation, Parity is still conserved
because the sign of the electric charge also changes:

\begin{equation}
\begin{array}{cccccccc}
P:& \ddot{\textbf{r}} &= &q(\textbf{E} + \dot{\textbf{r}} \times
\textbf{B})/m &\rightarrow & (-q)(\textbf{E} + \dot{\textbf{-r}}
\times \textbf{-B})/m &=& \ddot{\textbf{-r}}
\end{array}
\end{equation}

Classical electrodynamics cannot determine the transformation of
the physical quantities that appear in the equations. Under a parity operation
four possible options exist corresponding to scalar or pseudoscalar character for
the electric and/or magnetic charges. Two options conserve parity and two violate parity.
For aesthetic reasons we may favor the options that conserve parity. However
the strongest reason for parity conservation in electromagnetism is that
it is a consequence of a unified mathematical representation of
the fields (\ref{eq:F}) - that gives two possible options both of which conserve parity.

\section{A conserved Parity operator in Weak Interactions}
Returning to particle physics. The definition of a Parity
transformation has to include the transformations of mass, m,
Lepton numbers, $L_i$, Baryon number, $B$, Strangeness, $s$,
charm, $c$, top, $t$, and bottom, $b$. The origin of all these
quantum numbers is unknown and hence the appropriate
transformation properties can only be guessed at. The convention
is to define an operator which has no effect on all these
properties. Mathematically, these quantum numbers are
represented by scalars. The result is Parity violation in the Weak Interactions
and the puzzle of an asymmetric Universe.

Text books that introduce the Parity operator treat it as a
uniquely defined operator without mentioning the arbitrary choices
that have been made (see for example~\cite{Rolnick,Branco}). The experimental
evidence requires a more careful interpretation than is received
in most texts. Ballentine~\cite{ballentine} is one of the
few authors who notes that the inference of parity violation from
experimental evidence requires additional non-trivial assumptions
about the symmetry of the initial state - assumptions which he justifies.

An alternative definition of the Parity operator is to have
$m$ unchanged as before but all the other quantum numbers, $L_i,
B,s,c,t,b $ negated (mathematically they are pseudo-scalars)
and $q$ is also negated. Note that this definition requires the
electric rather than the magnetic field to be represented by an
axial vector, which is unconventional but consistent with
electrodynamics as described above. Within the Weinberg and Salaam
model of the weak interaction, such an operator is
well-known. It is commonly called $CP$. It is conserved by the
weak and strong interactions. If this is recognized as being the
appropriate operator for space inversion then the mystery of
left-right asymmetry in Weak Interactions is removed.

The requirement for Parity conservation in the model of the
weak interactions also leads to a unique parity conserving
operator for the electrodynamics. The Electric and magnetic
fields must have transformation properties opposite to the
conventional assignments so that magnetic fields and not
electric fields change sign under the Parity operator.

Without a deeper model of the elementary particles and their
quantum numbers, the assignment of their transformation properties
is only a matter of aesthetics. However the motivation for this
analysis came from just such models. The author is working on
geometric models of elementary particles and quantum theory
using classical general relativity \cite{hadley97,hadley98}. Some exciting results
have been achieved (see for example \cite{hadley2000}), but a clear prediction of such models
is that parity is conserved. General relativity conserves
parity. For every structure that displays handedness the
opposite structure is also a valid solution. General relativity
combined with symmetric boundary conditions must lead to a
parity conserving model of elementary particles. Far from
being a contradiction with experiment the analysis throws
new light on the definition of parity and challenges the
conventions that are universally and uncritically adopted.

\section{Conclusion}
The operator corresponding to space inversion is not unique. Making different, but equally valid assumptions leads to a Parity operator conserved in the weak and electromagnetic interactions. It is the operator normally denoted $CP$.

When making calculations using the weak interactions the new
identification for the Parity operator is irrelevant and inconsequential. But the Parity
operator and the supposed violation of Parity are used far more
widely (\cite{Hawking, Visser} are examples). The correct choice
of operator is vital to our understanding of Nature. Countless
textbooks published each year make statements about the asymmetry
of Nature based on the assumption that the conventional parity
operator is the unique representation of a space inversion.
Philosophers are left puzzling over the meaning of the asymmetry
and theoreticians looking for a unifying theory and have the
doubly difficult task of creating an asymmetric theory and
explaining why the mirror image theory is not seen.

Accepting that
a few quantum numbers can be pseudo-scalars rather than scalars is
sufficient to restore symmetry.

The violation of $CP$ in the neutral Kaon system remains
perplexing. But with the new definition of Parity, it is clear that the Universe itself has left-right asymmetry due to the predominance of positive Lepton and Baryon number. While such a fact is a long way short of being an explanation, the observation of an asymmetry in the presence of asymmetric boundary conditions is rather less surprising than Parity violation in a symmetric Universe.

\end{document}